\documentclass[preprint,showpacs,preprintnumbers,
amsmath,showkeys,amssymb,aps,floatfix]{revtex4}

\usepackage{epstopdf}
\usepackage{graphics}
\usepackage{graphicx}
\usepackage{dcolumn}
\usepackage{bm}
\usepackage{longtable}
\usepackage{epsfig}
\usepackage{times}
\usepackage{url}
\usepackage{color}

\begin{document}

\title{Nuclear Level Densities in the Constant--Spacing Model}

\author{Adriana P\'alffy}
\email{palffy@mpi-hd.mpg.de}

\author{Hans A. Weidenm\"uller}
\email{haw@mpi-hd.mpg.de}

\affiliation{Max-Planck-Institut f\"ur Kernphysik, Saupfercheckweg 1, D-69117 
Heidelberg, Germany}
\date{\today}
\begin{abstract}
 A new method to calculate level densities for non--interacting
  Fermions within the constant--spacing model with a finite number of
  states is developed. We show that asymptotically (for large numbers
  of particles or holes) the densities have Gaussian form. We improve
  on the Gaussian distribution by using analytical expressions for
  moments higher than the second. Comparison with numerical results
  shows that the resulting sixth--moment approximation is excellent
  except near the boundaries of the spectra and works globally for all
  particle/hole numbers and all excitation energies.

\end{abstract}
\maketitle

\section{Purpose}

Our interest in the dependence of various nuclear level densities on
energy and particle number has been triggered by recent experimental
developments in laser physics. The Extreme Light Infrastructure
(ELI)~\cite{eli} will open new possibilities for extremely
high--intensity laser interactions with fundamental quantum systems
different from the traditionally considered atoms and
molecules~\cite{Di12}. At the Nuclear Physics Pillar of ELI under
construction in Romania, efforts are under way to generate a
multi--MeV zeptosecond pulsed laser beam~\cite{Mor11}. For
medium--weight and heavy target nuclei interacting with such a beam,
photon coherence can cause multiple photon absorption. With energies of
several MeV per photon, the ensuing nuclear excitation energies may
well amount to several $100$ MeV. Depending on the time scale on which
the excitation takes place and on the specific nucleon-nucleon
interaction rates, collective excitations may be induced, or a
compound nucleus be formed~\cite{Wei11}. A theoretical treatment of
the latter process along the lines of precompound reaction models
requires the knowledge of the total level density, of the densities of
$p$--particle $h$--hole states, and of the density of accessible
states for particle/hole numbers and/or excitation energies that go
far beyond what has been considered until now. That applies not only
to the target nucleus but also to all daughter nuclei populated by
induced particle emission during the interaction time of the laser
pulse.

The standard approach to level densities goes back to the pioneering work of
Bethe~\cite{Bet36} who calculated the total level density as a
function of excitation energy with the help of the Darwin--Fowler
method. Basically the same method was used in many of the later
works~\cite{Eri60,Boe70,Wil71,Sta85,Obl86,Her92} dealing with the
density of $p$--particle $h$--hole states and related quantities. A
beautiful review is given in Ref.~\cite{Blo68}. The Darwin--Fowler
method yields analytical expressions involving contour integrals.
Their evaluation, although straightforward, becomes increasingly
involved with increasing numbers of particles and holes and/or
increasing excitation energy. The same is true for Refs.~\cite{Gho83,
 Bli86} that account for the exclusion principle by explicit
counting. Moreover, without explicit numerical calculation it is not
possible within these approaches to establish general properties of
particle--hole densities like the overall dependence on excitation
energy and/or particle--hole number. More recent works use a
static--path approximation (Refs.~\cite{Lau89, Alh93} and papers cited
therein) or account, in addition, for the residual interaction in an
approximate way (Ref.~\cite{Can94} and references therein). 
The method of Ref.~\cite{Hil98} avoids contour integrals and determines
(again numerically) the level densities directly as coefficients of
polynomials. The order of these rises rapidly, too, with energy and
particle/hole number. In none of
these approaches does it seem possible to deal with the enormously
large values of the various densities attained for medium--weight and
heavy nuclei at excitation energies of several $100$ MeV in a
practicable way. That is why we develop a different approach in the
present work.

In this Letter we  present
an analytical approximation to the global dependence of partial and
total level densities that takes full account of the exclusion
principle, that is valid for a finite number of single--particle
states, and that holds for all excitation energies and particle/hole
numbers. We prove analytically that the level density for particles or
holes is for a constant--spacing model asymptotically Gaussian. We improve on the Gaussian using
analytical results for the low moments of the distribution higher than
the second. Comparison with numerical results shows that the resulting
sixth--moment approximation is very precise except near the boundaries
of the spectrum (where numerical evaluation is easy). Particle--hole
densities follow by convolution. The attained analytical form of the
global dependence of level densities on excitation energy and particle
number extends our understanding of characteristic nuclear properties
into uncharted territory. Moreover, we expect our results to be an
indispensable tool in the calculation of laser--induced nuclear
reactions mentioned in the first paragraph.

We calculate the various densities in the framework of a
constant--spacing model for spinless non--interacting Fermions. To
justify our choice we consider by way of example  the partial level density $\rho_p(E, J,
\pi)$ for $p$ particles and $p$ holes, a function of excitation energy
$E$, total spin $J$, and parity $\pi$, for a system of
non--interacting Fermions in three dimensions. For other densities
the reasoning is the same. The partial level density is given by~\cite{Her92}
\begin{eqnarray}
\rho_p(E, J, \pi) &=& (1/2) \rho_p(E) \ \frac{2 J + 1}{2 \sqrt{2
\pi} \ \sigma^3_{2 p}}  \nonumber  \\
&\times&
\exp \bigg \{ - \frac{[J + (1/2)]^2}{2
\sigma^2_{2 p}} \bigg\} \ .
\label{0}
\end{eqnarray}
The factor $1/2$ accounts for parity. The last two terms of the
product give the spin dependence, with $\sigma_{2 p}$ the spin--cutoff
factor. With spin and parity being accounted for, $\rho_p(E)$ is
defined as the level density of spinless non--interacting Fermions
that carry no angular momentum. We note that in preequilibrium theories, the 
interactions between Fermions neglected here are taken into account as
agents for equilibration. In our model, the non--interacting Fermions are distributed over a set
of single--particle states. Each subshell with spin $j$ of the three-dimensional shell
model contributes $(2 j + 1)$ states to the set. For large excitation
energy or particle--hole numbers, we must take into account the
exclusion principle exactly. It is equally important to account for
the finite binding energy of particles and for the finite size of the
energy interval available for holes. Both strongly affect the various
level densities at large excitation energies. We do so using a
single--particle model with a finite number of states. Moreover, we
calculate the various densities using a constant--spacing model for
the single--particle states. It is clear from the shell model that the
model is not realistic at the high excitation energies of interest. Taking
into account the multiplicity $(2 j + 1)$ of the subshells, we note
that the single--particle level density of the shell model strongly
increases with energy. We return to this point at the end of Section \ref{numres}.

\section{Approach}

We consider $f$ spinless Fermions in a single--particle model with
constant level spacing $d$ and with a finite number $u$ of bound
single--particle states. In the ground state all single--particle
states from the lowest (energy $d$) up to a maximum level (energy $F =
f d$ with $F$ for Fermi energy) are occupied. The remaining $b = B /
d$ levels (with $B$ for binding energy) are empty. Here $f$ and $b$
are integers, and we have $u = f + b$. Excited states are described as
$p$--particle $h$--hole states, with $p$ counting the number of
particles in single--particle states with energy larger than $F$ and
not larger than $B + F$, and correspondingly $h$ counting the number
of holes with energy less than $F$. For non--closed shell compound
nuclei and/or nuclear reactions induced by composite particles, the
number of hole states $h$ may differ from $p$. We calculate various
many--body level densities for non--interacting particles: $\rho_B(p,
E)$ is the level density versus energy $E$ for $p$ particles confined
to an energy interval of length $B$, $\rho_F(h, E)$ is the level
density for $h$ holes confined to an energy interval of length $F$,
$\rho_{F B}(p, h, E)$ is the particle--hole state density defined
analogously, and $\rho_U(A, E)$ is the total level density for $A$
particles distributed over an energy interval of length $U = F +
B$. With $\varepsilon = E / d$ and $\varepsilon$ integer we define the
dimensionless density $\omega_b(p, \varepsilon) = \rho_B(p, E) \, d$
and analogously for $\omega_f(h, \varepsilon)$, $\omega_{f b}(p, h,
\varepsilon)$, and $\omega_u(A, \varepsilon)$. All densities denoted
by $\omega$ are integers.

We describe the method of calculation for $\omega_b(p, \varepsilon)$,
assuming for simplicity of notation that $b$ is odd and shifting the
energy such that the ground state of the $p$--particle system has
energy $(1/2) p (p + 1)$. The maximum energy is $b p - (1/2) p (p -
1)$, and the center of the spectrum is at
\begin{equation}
\varepsilon^{(0)}_b(p) = \frac{1}{2} p (b + 1) \ .
\label{1}
\end{equation}
The level density $\omega_b(p, \varepsilon)$ is defined as the number
of ways in which $p$ Fermions can be distributed over the $b$
available single--particle states such that the total energy equals
$\varepsilon$, i.e.,
\begin{equation}
\omega_b(p, \varepsilon) = \sum_{1 \leq n_1 < n_2 < \ldots < n_p
\leq b} \delta_{n_1 + n_2 + \ldots + n_p, \, \varepsilon} \ .
\label{2}
\end{equation}
The calculation of $\omega_b(p, \varepsilon)$ poses a purely
combinatorial problem. With $\beta = (1/2) (b - 1)$ we define new
summation variables $k_l = n_l - (1/2) (b + 1)$, $l = 1, 2, \ldots, p$
that range from $- \beta$ to $+ \beta$. With $\varepsilon' =
\varepsilon - \varepsilon_0(p)$ that gives
\begin{equation}
\omega_b(p, \varepsilon') = \sum_{-\beta \leq k_1 < k_2 < \ldots <
k_p \leq \beta} \delta_{k_1 + k_2 + \ldots + k_p, \, \varepsilon'} \ .
\label{3}
\end{equation}
We determine $\omega_b(p, \varepsilon')$ in terms of its low
moments. Changing the signs of all summation variables in
Eq.~(\ref{3}) one can easily show that $\omega_b(p, \varepsilon') =
\omega_b(p, - \varepsilon')$ is even in $\varepsilon'$, so that all
odd moments vanish. For the $2 m$th moment with $m = 0, 1, 2, \ldots$
we have
\begin{eqnarray}
m_b(p, 2 m) &=& \sum_{\varepsilon'} (\varepsilon')^{2 m} \omega_b(p,
\varepsilon') \nonumber \\
 &= & \sum_{- \beta \leq k_1 < k_2 < \ldots < k_p \leq \beta} \ \bigg(
\sum_l k_l \bigg)^{2 m} \ .
\label{4}
\end{eqnarray}
Following Ref.~\cite{Boe70} we adopt an occupation--number
representation for Fermionic many--body states. We represent each set
$\{ k_l \}$ of integers in Eq.~(\ref{3}) as a $b$--dimensional vector
$\{ \nu_1, \nu_2, \ldots, \nu_b \}$ with entries $\nu_j$ that take
values zero and one. The set $\{ k_l \}$ is represented by choosing
$\nu_j = 1$ in the $p$ positions $k_l$ and zero otherwise. The sum
over all $\{ k_l \}$ is replaced by the sum over all $b$--dimensional
vectors, i.e., over all choices of $\nu_j$ subject to the constraint
$\sum_j \nu_j = p$. Thus,
\begin{eqnarray}
&&m_b(p, 2 m) = \sum_{\nu_1, \nu_2, \ldots, \nu_b} \delta_{p, \nu_1 +
\nu_2 + \ldots + \nu_b} \ \bigg( \sum_j j \nu_j \bigg)^{2 m}
\nonumber \\
&&= \frac{\partial^{2 m}} {\partial \sigma^{2 m}} \sum_{\nu_1, \nu_2,
\ldots, \nu_b} \delta_{p, \nu_1 + \nu_2 + \ldots + \nu_b} \exp \{
\sigma \sum_j j \nu_j \} \bigg|_{\sigma = 0} \ .
\label{5}
\end{eqnarray}
We multiply Eq.~(\ref{5}) with $\exp \{ p \alpha \}$, sum over $p$,
and carry out the summations over the $\nu_j$. This gives the
partition function
\begin{equation}
Z_b(\alpha, 2 m) = \frac{\partial^{2 m}}{\partial \sigma^{2 m}}
\prod_{j = - \beta}^\beta (1 + \exp \{ \alpha + \sigma j \})
\bigg|_{\sigma = 0} \ .
\label{6}
\end{equation}
The moment $m_b(p, 2 m)$ is the coefficient multiplying $\exp \{
\alpha p \}$ in an expansion of $Z_b(\alpha, 2 m)$ in powers of $\exp
\{ \alpha \}$. For $m = 0$ we find $m_b(p, 0)$ $= {b \choose p}$, the
correct result. For $m = 1, 2$ we obtain
\begin{eqnarray}
m_b(p, 2) &=& \bigg( \sum_{j = - \beta}^\beta j^2 \bigg){ b - 2
\choose p - 1 } \ , \nonumber \\
m_b(p, 4) &=&\bigg( \sum_{j = - \beta}^\beta j^4 \bigg) \bigg[{ b - 4
\choose p - 1 } - 4 { b - 4 \choose p - 2 } + { b - 4 \choose p - 3 }
\bigg] \nonumber \\
&& + 3 \bigg( \sum_{j = - \beta}^\beta j^2 \bigg)^2 { b - 4 \choose
p - 2 } \ .
\label{7}
\end{eqnarray}
From Eqs.~(\ref{7}) we obtain the normalized moments
\begin{equation}
M_b(p, 2m) = \frac{m_b(p, 2 m)}{m_b(p, 0)} \ .
\label{8}
\end{equation}

\section{Asymptotically Gaussian Distribution}

Eqs.~(\ref{7}) suggest that asymptotically ($b \gg 1$, $p \gg 1$)
$\omega_b(p, \varepsilon')$ approaches a Gaussian distribution. (Here,
with $1 \leq p \leq b$ and $b \gg 1$, we consider $p \gg 1$ equivalent
to $p \approx b / 2$. Particle--hole symmetry connects the cases $p
\approx b$ and $p \approx 1$). We recall that for a Gaussian
distribution, the normalized fourth moment (see Eq.~(\ref{8})) equals
three times the square of the normalized second moment. For $b \gg1$
and $p \gg 1$ that is exactly the relation implied by the values of
$m_b(p, 4)$ and $m_b(p, 2)$ in Eq.~(\ref{7}). Indeed, taken by itself
the last term in the expression for $m_b(p, 4)$ yields a value for
$M_b(p, 4)$ which for $b \gg 1$, $p \gg 1$ equals three times the
square of $M_b(p, 2)$. Moreover, the term proportional to $ \sum j^4 $
in the expression for $m_b(p, 4)$ is smaller by the factor $1/p$ than
the one proportional to $ (\sum j^2)^2 $. To show that $\omega_b(p,
\varepsilon')$ becomes asymptotically ($b \gg 1, p \gg 1$) Gaussian we
generalize the approach of Eqs.~(\ref{5}) to (\ref{7}) to all
moments. We define
\begin{equation}
G(\sigma) = \prod_{j = - \beta}^\beta (1 + \exp \{ \alpha + \sigma j
\}) = \exp \{ H(\sigma) \}
\label{9}
\end{equation}
and expand $H(\sigma)$ in a Taylor series around $\sigma = 0$. With
\begin{equation}
f(\alpha) = \frac{\exp \{\alpha \}}{1 + \exp \{ \alpha \}}
\label{10}
\end{equation}
and $f^{(n)}$ denoting the $n$th derivative of $f$, we have for $m
= 1, 2, \ldots$
\begin{equation}
\frac{\partial^m}{\partial \sigma^m} H(\sigma) \bigg|_{\sigma = 0} =
f^{(m - 1)} \sum_{j = - \beta}^\beta j^m \ .
\label{11}
\end{equation}
This shows that all odd derivatives of $H$ vanish. We insert the
Taylor expansion for $H$ into Eq.~(\ref{9}) and obtain
\begin{equation}
G(\sigma) = G(0) \exp \bigg\{ \sum_{n = 1}^\infty \frac{1}{(2 n)!}
\sigma^{2 n} f^{(2 n - 1)} \sum_{j = - \beta}^\beta j^{2 n} \bigg\}
\ .
\label{12}
\end{equation}
From here, we proceed in two steps. (i) We neglect all terms with $n >
1$ on the right--hand side of Eq.~(\ref{12}) and show that as a
result, $\omega_b$ is Gaussian for $b \gg 1$, $p \gg 1$. (ii) We show
by complete induction that all terms with $n > 1$ in Eq.~(\ref{12})
are negligibly small in the same limit. (i) For $\omega_b(p,
\varepsilon')$ to be Gaussian we have to show that $M_b(p, 2 m) = (2 m
- 1)!! \ [M_b(p, 2)]^m$. Taking into account the term with $n = 1$
only, expanding the exponential, and using the result in
Eqs.~(\ref{9}) and (\ref{6}) we obtain
\begin{equation}
Z_b(\alpha, 2 m) = (2 m - 1)!! \ G(0) (f')^m \bigg( \sum_{j = -
\beta}^\beta j^2 \bigg)^m \ .
\label{13}
\end{equation}
The normalized $2 m$th moment $M_b(p, 2 m)$ is the coefficient
multiplying $\exp \{ p \alpha \}$ in the expansion of $Z_b(\alpha, 2
m)$ in powers of $\exp \{ \alpha \}$ divided by the normalization
factor ${b \choose p}$. For $b \gg 1$ and $p \gg 1$ the relevant
coefficient in $G(0) (f')^m {b \choose p}^{- 1}$ is $\approx [p (b -
p) b^{- 2}]^m$.  That yields $M_b(p, 2 m) \approx (2 m - 1)!! \
[M_b(p, 2)]^m$, consistent with a Gaussian form for $\omega_b(p,
\varepsilon')$. In the last step of the argument we approximate
products of the form $p (p - 1) \ldots (p - m)$ by $p^m$. For fixed
$p$ the approximation becomes increasingly inaccurate as $m$
increases.  Our result is therefore valid only asymptotically. (ii) We
use complete induction to show that the contributions of the terms
with $n > 1$ in Eq.~(\ref{12}) become vanishingly small for $b \gg 1,
p \gg 1$. We have shown above that this claim holds for $n = 2$ (i.e.,
for $m_b(p, 4)$). We assume that the claim is correct for $2 \leq n <
n_0$, omit the corresponding terms in Eq.~(\ref{12}), and show that it
holds for $n = n_0$, i.e., for $M_b(p, 2 n_0)$. We have shown under
(i) that the contribution to $M_b(p, 2 n_0)$ of the term with $n = 1$
is $(2 n_0 - 1)!! \ (M_b(p, 2))^{n_0}$. The contribution of the term
with $n = n_0$ is $G(0) f^{(2 n_0 - 1)} \sum_j j^{2 n_0}$. For $b \gg
1$ we have $\sum_j j^m \approx b^{m + 1} / ( 2^m (m + 1))$. From
Eq.~(\ref{10}) we have $f' = f - f^2$. Therefore, $f^{(2 n_0 - 1)} =
\sum_{l = 1}^{2 n_0} c_l f^l$ is a polynomial of degree $2 n_0$ in $f$
with integer coefficients $c_l$, and the contribution of $G(0) f^{(2
  n_0 - 1)} \sum_j j^{2 n_0}$ to $M_b(p, 2 n_0)$ is
\begin{equation}
{ b \choose p }^{-1} \frac{b^{2 n_0 + 1}}{(2 n_0 + 1) 2^{2 n_0}}
\sum_{l = 1}^{2 n_0} c_l { b - 2 n_0 \choose p  - l} \ .
\label{14}
\end{equation}
For $b \gg 2 n_0$ and $p \gg l$ we have ${ b - 2 n_0 \choose p - l}
\approx { b \choose p } p^l / b^l$. The contribution~(\ref{14}) is,
therefore, of order $b^{2 n_0 + 1}$ while the contribution from the
term with $n = 1$ is of order $b^{3 n_0}$.  This shows that the
contributions with $n = 1$ dominate all others. The situation differs
for $b \gg 1$ and $p \approx 1$ or $p \approx b$ where $M_b(p, 2)$ is
of order $b^2$ only and $[M_b(p, 2)]^{n_0}$ is comparable in size to
the contribution~(\ref{14}). Here the Gaussian approximation cannot be
expected to work well. This is consistent with the fact that for $p =
1$ and $p = b - 1$ the densities are flat, $\omega_b(1, \varepsilon')
= 1 = \omega_b(b - 1, \varepsilon')$. Furthermore, for $p = 2$ and $p
= b - 2$ the densities have a triangularly shaped maximum. Only with
$p = 3$ and $p = b - 3$ does the density of states become
Gaussian--shaped. The maximum at $\varepsilon' =
\varepsilon^{(0)}_b(p)$ builds up only slowly as $p$ increases from
unity or decreases from $b - 1$.

\section{Low-Moments Approximation}

Using the asymptotically Gaussian form of $\omega_b(p, \varepsilon')$
we approximate that function in terms of its low even moments. We use
Eqs.~(\ref{7}) for $m_b(p, 2)$ and $m_b(p, 4)$, calculate $m_b(p, 6)$
similarly, and find the parameters $\gamma_{2 m}$, $m = 1, 2, 3$ of
the normalized function
\begin{equation}
F^{(6)}_b(p, \varepsilon') = C \exp \{ - \gamma_2 (\varepsilon')^2 -
\gamma_4 (\varepsilon')^4 - \gamma_6 (\varepsilon')^6 \}
\label{15}
\end{equation}
that correspond to the normalized moments $M_b(p, 2 m)$ with $m = 1,
2, 3$ in Eq.~(\ref{8}). The resulting function
\begin{equation}
\omega^{(6)}_b(p, \varepsilon') = {b \choose p} F^{(6)}_b(p,
\varepsilon')
\label{16}
\end{equation}
is referred to in the following as the sixth--moment approximation to
$\omega_b(p, \varepsilon')$. Approximations obtained by using only the
second (only the second and the fourth) moment(s) are denoted by
$\omega^{(2)}_b(p, \varepsilon')$ (by $\omega^{(4)}_b(p,
\varepsilon')$, respectively). The same approach is used for
$\omega_f(h, \varepsilon')$ and for $\omega_u(A, \varepsilon)$. Except
for suitable changes of indices and parameters, the results are
formally identical. We mention in passing that the method is also
useful for calculating the density of accessible states~\cite{Obl86,
Car12} under the constraints of the exclusion principle. Details
will be given elsewhere. For the $p$--particle $h$--hole density
$\omega_{b f}(p, h, \varepsilon)$ we define $\varepsilon$ as the total
excitation energy of the Fermionic system. Then
\begin{equation}
\omega_{b f}(p, h, \varepsilon) = \sum_{\varepsilon_p \varepsilon_h}
\delta_{\varepsilon_p + \varepsilon_h, \,
\varepsilon} \omega_b(p, \varepsilon_p) \omega_f(h, \varepsilon_h) \ .
\label{17}
\end{equation}
Here $\varepsilon_p = \varepsilon^{(0)}(p) + \varepsilon'_p$ is the
total energy of the $p$ particles, and correspondingly for holes,
while $\varepsilon = \varepsilon_p + \varepsilon_h$ is the excitation
energy of the $p$--particle $h$--hole system. Thus
\begin{equation}
\omega_{b f}(p, h, \varepsilon) = \sum_{\varepsilon_p \varepsilon_h}
\delta_{\varepsilon'_p + \varepsilon'_h,\, \varepsilon -
\varepsilon^{(0)}_p - \varepsilon^{(0)}_h} \omega_b(p, \varepsilon'_p)
\omega_f(h, \varepsilon'_h) \ .
\label{18}
\end{equation}
Since $\omega_b(p, \varepsilon'_p)$ ($\omega_f(h, \varepsilon'_h)$) is
a symmetric function of $\varepsilon'_p$ (of $\varepsilon'_h$,
respectively), it follows that $\omega_{b f}$ is a symmetric function
of $\varepsilon$ centered at $\varepsilon^{(0)} = \varepsilon^{(0)}_p
+ \varepsilon^{(0)}_h$. Therefore we consider the function $\omega_{b
f}(p, h, \varepsilon')$ with $\varepsilon' = \varepsilon -
\varepsilon^{(0)}$. This function is symmetric about $\varepsilon' =
0$. For the low even moments we obtain
\begin{eqnarray}
m_{b f}(p, h, 0) &=& {b \choose p} \ {f \choose h} \ , \nonumber \\  
m_{b f}(p, h, 2) &=& m_b(p, 2) + m_f(h, 2) \ , \nonumber \\
m_{b f}(p, h, 4) &=& m_b(p, 4) + 2 m_b(p, 2) m_f(h, 2) 
\nonumber \\
&&+ m_f(h, 4) \ ,
\label{19}
\end{eqnarray}
and correspondingly for higher moments.

\section{Numerical Results \label{numres}}

\begin{figure}[ht]
\vspace{5 mm}
\includegraphics[width=0.8\linewidth]{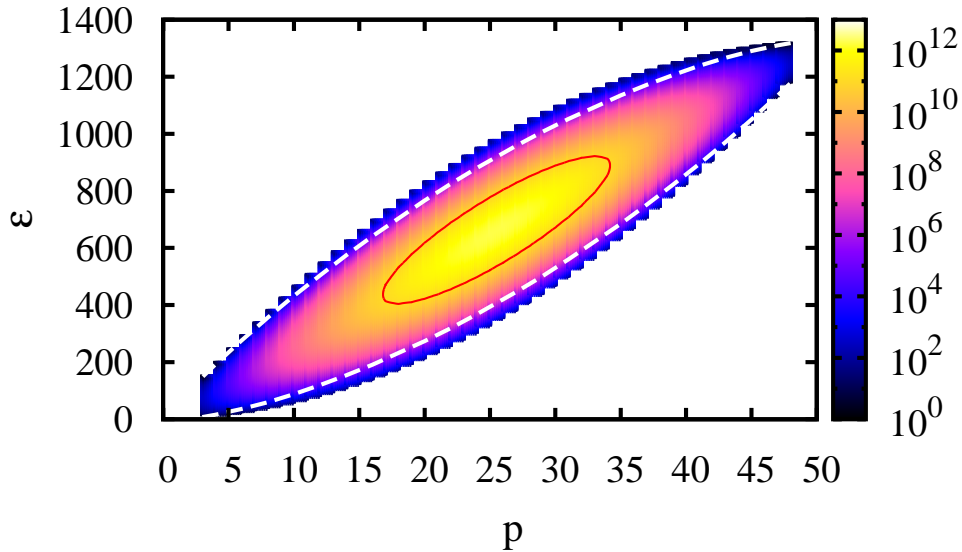}
\vspace{3 mm}
\caption{Contour plot of the dimensionless level density $\omega^{(6)}_b(p,
\varepsilon)$ for $p$ particles in $b= 51$ equally spaced single--particle
states based on the sixth--moment approximation~(\ref{16}) versus $p$
and versus dimensionless energy $\varepsilon$. The white dashed lines
define the boundary of the region where $\omega^{(6)}_b(p,
\varepsilon)$ deviates by $10\%$ or more from the exact values. The full
line presents the constant contour $\omega^{(6)}_b(p,
\varepsilon)=10^{11}$. }
\label{fig1}
\end{figure}

We begin with an overview of the dependence of $\omega_b(p,
\varepsilon)$ on both $p$ and $\varepsilon$ using the sixth--moment
approximation~(\ref{16}). Even though we expect that approximation to
work well only for $p \approx b / 2$, we display in Fig.~\ref{fig1}
the values of $\omega^{(6)}_b(p, \varepsilon)$ for $b = 51$ in the
$p$--$\varepsilon$ plane as a coloured contour plot for all values of
$p$ between 3 and $b - 3$. For fixed $p$, the dimensionless energy
$\varepsilon$ takes values in the interval $(1/2) p (p + 1) \leq
\varepsilon \leq b p - (1/2) p (p - 1)$. This accounts for the two
nearly parabolic and sawtooth--like boundaries of the coloured
domain. The parabolic dependence on $p$ is given by $(1/2) p (p + 1)$
for the lower edge and by $b p - (1/2) p^2 + (1/2) p$ for the upper
edge. The contour plot is symmetric with respect to a simultaneous
mirror reflection about the vertical line $p = (b - 1)/ 2$ and about
the horizontal line defined by the overall centroid energy
$\varepsilon = (1/4) (b^2 - 1)$. This symmetry is due to the symmetry
of $\omega_b(p, \varepsilon)$ in $\varepsilon$ about the centroid
energy $\varepsilon^{(0)}_b(p)$, and to particle--hole symmetry which
equates $\omega_b(p, \varepsilon)$ with $\omega_b(b - p, \varepsilon)$
except for a shift by the difference $\varepsilon^{(0)}_b(b - p) -
\varepsilon^{(0)}_b(p)$ of the centroid energies. For fixed $p$,
$\omega_b(p, \varepsilon)$ displays a maximum at
$\varepsilon^{(0)}_b(p) = (1/2) p (b + 1)$ (except for the cases $p =
1$ and $p = b - 1$ not displayed in the Figure). The location of the
maximum increases linearly with $p$. This fact and the parabolic form
of the boundaries cause the quasi--elliptical shape of the solid line
of constant $\omega_b$--values in the colour plot. We note the
enormous maximum values of $\omega_b(p, \varepsilon^{(0)}_b(p))
\approx 10^{12}$ attained for $p \approx 25$. All these features are
generic (i.e., independent of the performance of the sixth--moment
approximation) and apply likewise to $\omega_f(h, \varepsilon')$ and
to $\omega_u(A, \varepsilon')$, except for a rescaling of abscissa,
ordinate, and of the values of the densities.

Limitations of the sixth--moment approximation become obvious when we
consider the values of $\omega_b(p, \varepsilon)$ at the boundaries
$\varepsilon = (1/2) p (p + 1)$ and $\varepsilon = b p - (1/2) p (p -
1)$ where we obviously must have $\omega_b(p, \varepsilon) \approx
1$. The sixth--moment approximation exceeds this value by one to two
orders of magnitude, see Fig.~\ref{fig3} below. (We should keep in
mind, of course, that the values at the boundaries predicted by the
sixth--moment approximation are smaller by about $10$ orders of
magnitude than the values in the maximum. The relative accuracy of the
sixth--moment approximation is, therefore, excellent). The white
dashed lines in Fig.~\ref{fig1} show at which values of $p$ and
$\varepsilon$ the sixth--moment approximation deviates by $10 \%$ from
the exact values. The inaccuracy affects only the very tails of the
density of states, symmetrically about the centroid energy
$\varepsilon^{(0)}_b(p)$. The dashed lines thus follow the boundaries
of $\omega_b(p,\varepsilon)$.

We test the performance of the sixth--moment approximation in detail
by a comparison with exact numerical results. This can be done
throughout the critical domain (where either $p \approx 1$ or $p
\approx b$ or where $\varepsilon$ is close to the boundary of the
spectrum) since this is easily accessible numerically. We calculate
the exact values of $\omega_b(p, \varepsilon)$ in two ways. (i) We
directly use Eq.~(\ref{3}). (ii) We use the occupation--number
representation defined above Eq.~(\ref{5}) and sum over all
$b$--dimensional vectors $\{\nu_1, \nu_2, \ldots, \nu_b \}$, grouping
the results according to particle number $p = \sum_j \nu_j$ and energy
$\varepsilon' = \sum_j j \nu_j$. Method (i) yields $\omega_b(p,
\varepsilon)$ for fixed $b$ and $p$. Method (ii) yields $\omega_b(p,
\varepsilon')$ for fixed $b$ and all values of $p$ and
$\varepsilon'$. The demand on computing time is obviously larger for
method (ii). Particle--hole densities are then obtained from
Eq.~(\ref{17}). Our results agree with those of Ref.~\cite{Obl86} for
the small numbers of particles and holes considered there.

\begin{figure}[ht]
\vspace{5 mm}
\includegraphics[width=0.8\linewidth]{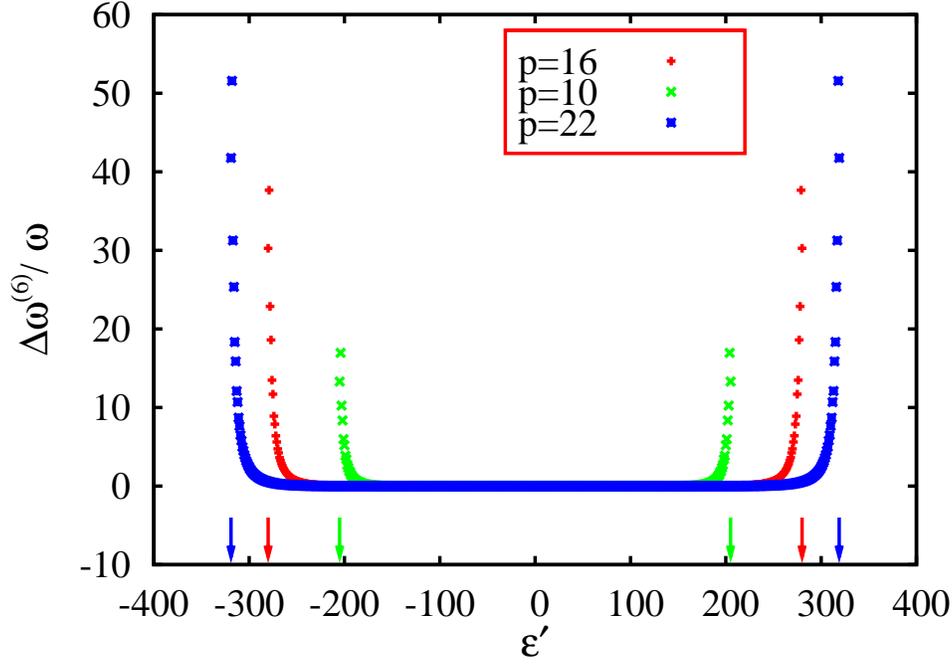}
\vspace{3 mm}
\caption{Relative difference between the exact result and the
sixth--moment approximation
$\Delta\omega^{(6)}/\omega=[\omega^{(6)}_b(p,
\varepsilon')-\omega_b(p, \varepsilon')]/\omega_b(p, \varepsilon')$
versus energy $\varepsilon'$ for $b = 51$ and several values of
$p$. The spectral boundaries are indicated by the arrows on the
abscissa.}
\label{fig2}
\end{figure}

For the comparison between our exact and approximate results, we
restrict ourselves to a few central features. In Fig.~\ref{fig2} we
display the relative difference between the exact result and the
sixth--moment approximation for $b = 51$ and various values of $p$
near $b / 2$. Significant deviations occur only at the boundaries of
the spectrum where the values of the sixth--moment approximation are
too large. Even though the tails of the sixth--moment approximation
are suppressed by many orders of magnitude in comparison with the
value at the center, that suppression is not strong enough. This is
shown more clearly in Fig.~\ref{fig3} where we plot for $b = 51$ the
values of $\omega^{(2)}_b(p, \varepsilon)$, of $\omega^{(4)}_b(p,
\varepsilon)$, and of $\omega^{(6)}_b(p, \varepsilon)$ (these
functions are defined in and below Eq.~(\ref{16})) versus $p$ at the
boundary of the spectrum for $b = 51$.  With every additional moment
included in the approximation, the agreement with the correct value
$\omega_b \approx 1$ is drastically improved. However, it would
obviously take even higher moments than the sixth one to reach
quantitative agreement at the boundary of the spectrum. This can be
done, although convergence may be slow. Alternatively, we may
calculate $\omega_b(p, \varepsilon)$ numerically for the critical
values of $p$ and $\varepsilon$ where the sixth--moment approximation
is not sufficiently precise. These values lie at the boundaries of the
spectrum shown in Figure~\ref{fig1} where either $p \approx 1$ or $p
\approx b$ or $\varepsilon \approx (1/2) p (p + 1)$ or $\varepsilon
\approx p b - (1/2) p (p - 1)$. In all these cases the number of terms
that contribute to Eq.~(\ref{3}) is small, and the calculation is
straightforward. Furthermore, for energies close to the spectrum
boundaries, the density of states depends only on $p$ and not on the
number of levels $b$. For the lower boundary that is the case for
$(1/2) p (p + 1) \leq \varepsilon \leq (1/2) p (p - 1) + b + 1$. The
numerical calculation can then be performed conveniently for a smaller
number of particle states $b'$ chosen such that $\omega_b(p,
\varepsilon) = \omega_{b'}(p, \varepsilon)$ for $\varepsilon$ in the
energy interval of interest.  For the dashed lines in Fig.~\ref{fig1}
defining a $10\%$ deviation of our approximate results, for instance,
the exact values can be calculated numerically using $b'\simeq
p(b-p)/10+p-1$. It should also be borne in mind that in preequilibrium
calculations one typically requires ratios (and not absolute values)
of densities. We expect that these are predicted quite precisely by
the sixth--moment approximation even at the boundaries of the
spectrum.

\begin{figure}[ht]
\vspace{5 mm}
\includegraphics[width=0.8\linewidth]{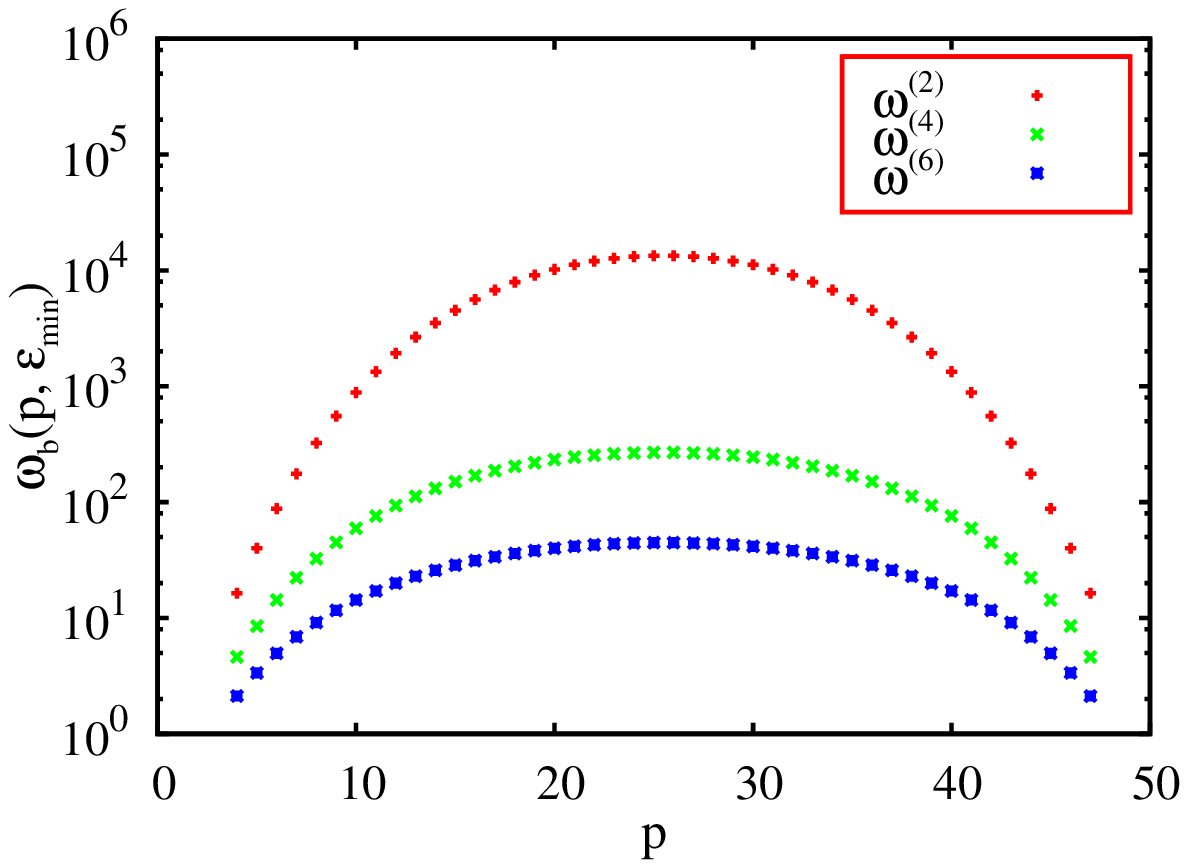}
\vspace{3 mm}
\caption{ Second--,  fourth--, and  sixth--moment approximation to
$\omega_b(p, \varepsilon)$  at the lower spectrum boundary
$\varepsilon_{\mathrm{min}} = (1/2) p (p + 1)$ versus $p$ for
$b = 51$.}
\label{fig3}
\end{figure}

\begin{figure}[ht]
\vspace{5 mm}
\includegraphics[width=0.8\linewidth]{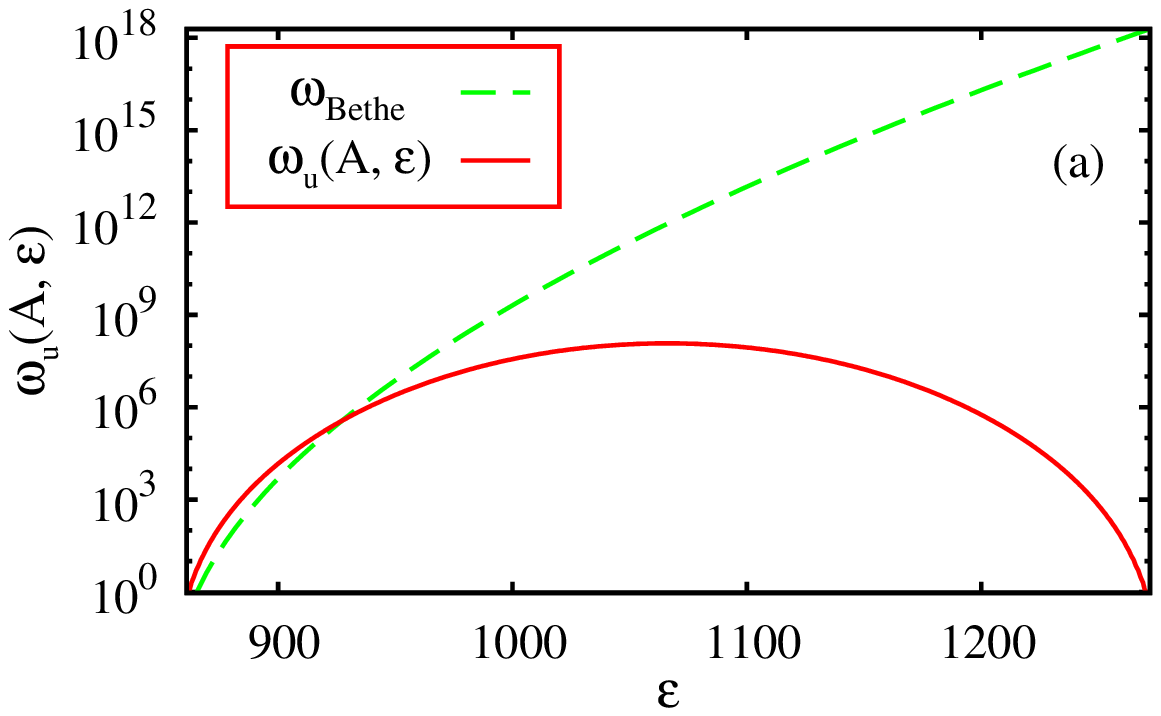}
\includegraphics[width=0.8\linewidth]{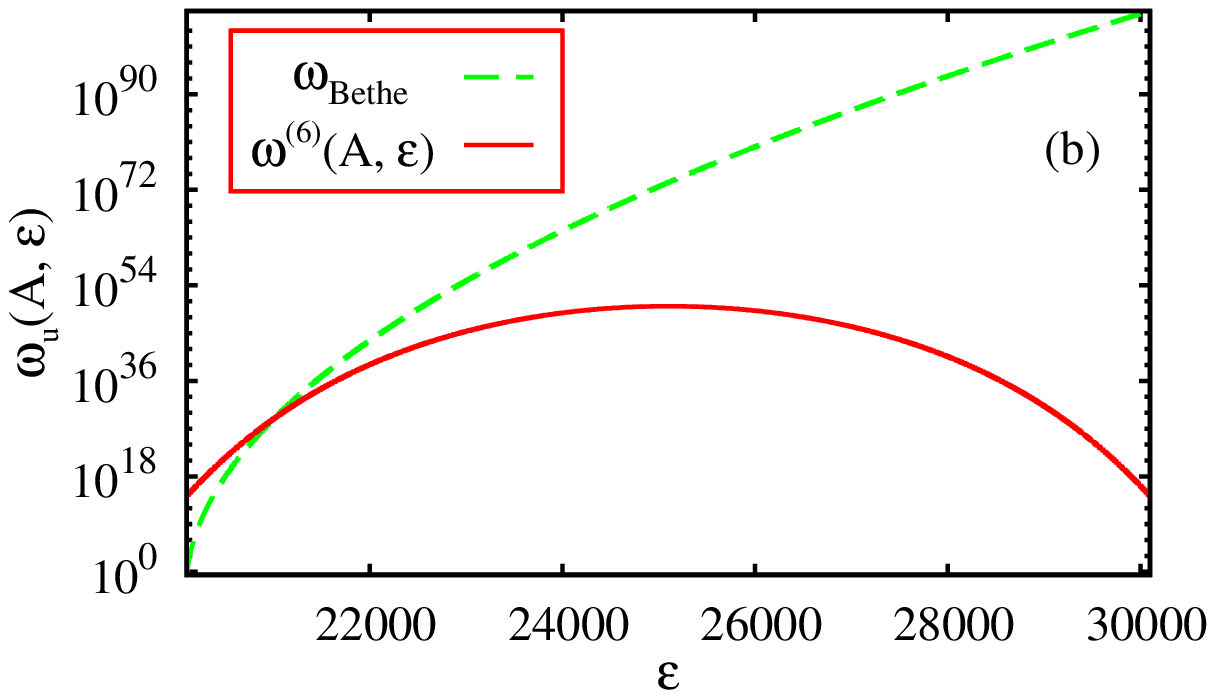}
\vspace{3 mm}
\caption{(a) Comparison of the exact level density $\omega_u(A,
\varepsilon)$ for $u = 51$, $A = 41$ versus energy $\varepsilon$ with
the Bethe formula~(\ref{20}). (b) The same for the sixth--moment
approximation and $u = 250$, $A = 200$. }
\label{fig4}
\end{figure}

We turn to the total level density $\omega_u(A, \varepsilon)$ of $A$
particles distributed over $u$ equally spaced single--particle states
as a function of the dimensionless excitation energy $\varepsilon$. We
have shown that $\omega_u(A, \varepsilon)$ has approximately Gaussian
shape, with a peak at half the total excitation energy $(1/2) A (u +
1)$. The original calculation of $\omega_u(A, \varepsilon)$ by
Bethe~\cite{Bet36} effectively also used a constant--spacing model but
neglected the limitations due to finite particle number $A$ and finite
number $u$ of single--particle states. With energy measured in units
of the single--particle level spacing, the celebrated ``Bethe
formula'' reads~\cite{Bet36}
\begin{equation}
\omega_{\rm Bethe}(\varepsilon) = \frac{1}{\varepsilon \sqrt{48}} \exp
\{ \pi \sqrt{2 \varepsilon / 3} \} \ .
\label{20}
\end{equation}
We note that $\omega_{\rm Bethe}(\varepsilon)$ does not contain any
adjustable parameters. The singularity at $\varepsilon = 0$ is due to
the Darwin--Fowler method. The ensuing approximation fails at and near
$\varepsilon = 0$. Beyond this domain the Bethe formula yields a
monotonically rising function of excitation energy since the
underlying counting method assumes that the number of available
single--particle states is unbounded. Thus, there exists an excitation
energy beyond which the Bethe formula exceeds $\omega_u(A,
\varepsilon)$ by an ever growing amount. This fact was qualitatively
pointed out in Ref.~\cite{Wei64}. In Fig.~\ref{fig4} we compare
$\omega_{\rm Bethe}(\varepsilon)$ with the exact calculation for $u =
51$ and $A = 41$ and with the sixth--moment approximation for $u =
250$ and $A = 200$ (for the latter parameters the exact density of
state values are not available for the entire energy spectrum). The
latter parameter set mimics, very roughly, a heavy nucleus.
Comparison with the exact calculation shows that $\omega_{\rm
  Bethe}(\varepsilon)$ underestimates the level density below the
crossing point of both curves. We have found this to be a systematic
trend. Both parts of Fig.~\ref{fig4} clearly display the crossing
point and the increasing discrepancy between $\omega_{\rm
  Bethe}(\varepsilon)$ and $\omega_u(A, \varepsilon)$ as $\varepsilon$
increases beyond this point.
We interpret the data on the crossing points using the equilibrium
distribution $n(\epsilon) = (1/A) \ 1/(1 + \exp \{ (\epsilon - A)/(kT)
\}$ for $A$ Fermions at temperature $T$ with $kT \ll A$ and continuous
single--particle energy $\epsilon$. With $kT \approx
\sqrt{\varepsilon}$ where $\varepsilon$ is the total excitation energy
of the many--body system, we find that the crossing points occur at an
excitation energy $\varepsilon$ where the fraction of particles in
states with energies $> u$ is of the order of a few percent. This is
physically plausible. In a heavy nucleus this criterion corresponds to
excitation energies around $200$ MeV. 

With increasing excitation energy, the constant--spacing model becomes
increasingly unrealistic. Indeed, the standard value~\cite{Her92} $d
\approx 13 / A$ MeV for the average spacing of single--particle levels
near the Fermi energy in medium--weight and heavy nuclei strongly
underestimates the single--particle level spacing in low--lying
shells. We recall that every subshell with spin $j$ contributes $(2 j
+ 1)$ states to $\rho_p(E)$ in Eq.~(\ref{0}). As a consequence, the number of states available
for high--energy hole formation is smaller than predicted by the
constant--spacing model. Therefore, the actual level density bends
over more strongly with increasing excitation energy and terminates at
a lower maximum energy than shown in Fig.~\ref{fig4}, and the
discrepancy with the Bethe formula is even bigger than presented
there. The effect of an energy--dependent single--particle level density 
was previously addressed, for instance, in Refs.~\cite{Bog92, Shl95} albeit under neglect of the
exclusion principle.  

To account for the shortcoming of the constant--spacing model we are in the process of
improving our approach to calculate nuclear level densities. We divide the energy interval
$U = B + F$ into several sections $l = 1, 2, \dots$ with constant
level spacing $d_l$ each but with $d_l \neq d_{l'}$ for $l \neq
l'$. Distributing $p$ particles in all possible ways over these
sections, so that there are $p_l$ particles in section $l$, we can use
our results for the constant--spacing model in each section
separately. For a fixed distribution $\{ p_l \}$ the level density is
a convolution over a product of Gaussians. The total level density is
the sum over all distributions $\{ p_l \}$. We note that it will no longer be a symmetric
function of energy.

\section{Conclusions}

Combining analytical and numerical methods, we have used a
constant--spacing model for non--interacting spinless Fermions to
develop a global approach to various nuclear level densities. This
approach is viable also for large particle numbers and/or excitation
energies where previous work runs into difficulties. As representative
example we have displayed in detail the calculation of the particle
density $\omega_b(p, \varepsilon)$ as a function of particle number
$p$ and excitation energy $\varepsilon$. This function is symmetric
about the center of the spectrum and, except for $p = 1$ and $p = b -
1$, displays a maximum at the center. It also possesses particle--hole
symmetry. The shape of the boundaries of the spectrum and the two
symmetries are responsible for the quasi--elliptical shape of the line
of constant density in Fig.~\ref{fig1}. With $\omega_b(p, \varepsilon)
\approx 1$ at the boundaries of the spectrum, the value of
$\omega_b(p, \varepsilon)$ at the center increases dramatically with
increasing $b$ and $p \approx b / 2$, reaching values near $10^{12}$
already for $b \approx 50$ (and even larger values as $b$ is further
increased). The decrease by $12$ or more orders of magnitude from the
center of the spectrum to the boundary poses a considerable challenge
to viable analytical approximations. Guided by the fact that for $b
\gg 1, p \gg 1$ $\omega_b(p, \varepsilon)$ becomes asymptotically a
Gaussian function of energy, we have used analytical expressions for
the low moments to determine a sixth--moment approximation to
$\omega_b(p, \varepsilon)$. This approximation is excellent except for
values of $p$ and $\varepsilon$ near the boundaries of the
spectrum. These are indicated by the white dashed lines in
Fig.~\ref{fig1}. Here the numerical calculation of $\omega_b(p,
\varepsilon)$ based on Eq.~(\ref{3}) is easy and fast.  Combining both
approaches we obtain a reliable and easy--to--handle method of
calculating the overall nuclear level density and $p$--particle
$h$--hole densities for medium--weight and heavy nuclei for all
particle numbers and at all excitation energies. The results should be
realistic except for limitations due to the underlying
constant--spacing model. Because of shell effects the density of
single--particle levels increases towards the Fermi energy, and this
fact is not taken into account in the model. Work on a suitable
generalization is under way. The Bethe formula is seen to fail beyond
an excitation energy that amounts to approximately $200$ MeV in heavy
nuclei.


\end{document}